\documentclass[]{jpsj3}
\usepackage{graphicx}% Include figure files
\usepackage{dcolumn}% Align table columns on decimal point
\usepackage{bm}% bold math

\title{Quantum Spin-Liquid Behavior in the Spin-1/2 Random Heisenberg Antiferromagnet on the Triangular Lattice}

\author{Ken Watanabe, Hikaru Kawamura\thanks{E-mail:kawamura@ess.sci.osaka-u.ac.jp}, Hiroki Nakano$^{1}$, and T\^oru Sakai$^{1,2}$}

\inst{Department of Earth and Space Science, Faculty of Science, Osaka University, Toyonaka 560-0043, Japan\\
$^{1}$Graduate School of Material Science, University of Hyogo, Kamigori, Hyogo 678-1297, Japan\\
$^{2}$Japan Atomic Energy Agency, SPring-8, Sayo, Hyogo 679-5148, Japan}

\def\gsim{\buildrel {\textstyle >}\over {_\sim}}
\def\lsim{\buildrel {\textstyle <}\over {_\sim}}

\abst{
Experimental quest for the hypothetical ``quantum spin liquid'' state has recently met a few promising candidate materials including organic salts $\kappa$-(ET)$_2$Cu$_2$(CN)$_3$ and EtMe$_3$Sb[Pd(dmit)$_2$]$_2$, $S=1/2$ triangular-lattice  Heisenberg antiferromagnets consisting of molecular dimers. These compounds exhibit no magnetic ordering nor the spin freezing down to very low temperature, while various physical quantities exhibit gapless behaviors. Recent dielectric measurements revealed the glassy dielectric response suggesting the random freezing of the electric polarization degrees of freedom. Inspired by this observation, we propose as a minimal model of the observed quantum spin-liquid behavior the $S=1/2$ antiferromagnetic Heisenberg on the triangular lattice with a quenched randomness in the exchange interaction. We study both zero- and finite-temperature properties of the model by an exact diagonalization method, to find that when the randomness exceeds a critic
 al value the model exhibits a quantum spin-liquid ground state. This randomness-induced quantum spin-liquid  state exhibits gapless behaviors including the temperature-linear specific heat. The results provide a consistent explanation of the recent experimental observations on organic salts. 
}

\kword{frustration, triangular lattice, spin liquid, randomness, random-singlet phase, valence-bond glass }

\begin{document}
\maketitle

\section{Introduction}

 The possibility of the so-called quantum spin-liquid state which accompanies no magnetic long-range order (LRO) down to low temperature, neither periodic nor random, has long attracted interest of researchers \cite{Levy,Lee-review,Balents}. This long-standing interest was largely motivated by the proposal by P.W. Anderson of the resonating valence bond (RVB) state possibly realized in the quantum $S=1/2$ Heisenberg antiferromagnet on the triangular lattice \cite{Anderson}. Since this proposal, much effort has been devoted to realize the quantum spin-liquid state in real magnetic materials. Only recently, there reported several candidate materials in certain frustrated magnets. In particular, it was reported that $S=1/2$ organic triangular-lattice salts $\kappa$-(ET)$_2$Cu$_2$(CN)$_3$ \cite{Shimizu,Kurosaki,Shimizu2,Ohira,SYamashita,MYamashita,Manna,Jawad,Pratt,Poirier,Sedlmeier,Itoh} and EtMe$_3$Sb[Pd(dmit)$_2$]$_2$ \cite{Itou,Itou2,MYamashita2,SYamashita2,Watanabe,Jawad2} e
 xhibited such a quantum spin-liquid-like behavior. In $\kappa$-(ET)$_2$Cu$_2$(CN)$_3$, the spin does not order down to 32mK which is four orders of magnitudes lower than the Curie-Weiss temperature of this compound $\sim 250$K \cite{Shimizu}. Interestingly, the low-temperature specific heat \cite{SYamashita,SYamashita2} or the thermal conductivity \cite{MYamashita,MYamashita2} exhibit a behavior linear in the absolute temperature $T$ (a small gap is observed in the thermal conductivity of the ET compound, though \cite{MYamashita}), which is rather unusual for insulating magnets.  Theoretical analysis revealed that a simple $S=1/2$ antiferromagnetic (AF) Heisenberg model with the nearest-neighbor bilinear interaction exhibited a N\'eel LRO with the three-sublattice periodicity \cite{Bernu,Capriotti}, the so-called 120-degrees structure. Given that, the spin-liquid state in organic compounds requires some ingredients not taken into account in the simplest Heisenberg model with
  the nearest-neighbor bilinear coupling. In spite of several theoretical proposals and numerical calculations \cite{LiMing,Morita,Motrunich,Lee,Yunoki,Lee2,Xu,Tocchio,Tocchio2}, the true nature of the observed spin-liquid state of these organic salts compounds has still remained elusive. 

 In these organic salts, the triangular lattice on which the spin-1/2 lies consists of molecular dimers over which the spin density of $S=1/2$ is spatially spread \cite{Jawad,Hotta,Naka,Dayal}. Hence, the electric-polarization degrees of freedom actually exist at each dimer site of the triangular lattice. An interesting recent experimental observation on $\kappa$-(ET)$_2$Cu$_2$(CN)$_3$ is that these dielectric degrees of freedom might be slowed down and tend to be frozen at lower temperatures. Indeed, the measured ac dielectric constant exhibited a glassy response in the temperature range of $T\lesssim 30K$ even on the macroscopic time scale of kHz, indicating the possible occurrence of the random freezing of the electric polarization degrees of freedom \cite{Jawad}. Then, in the temperature range of the quantum spin liquid (in the ET salt the $T$-linear specific heat is observed at $T\lesssim 6K$), the bias in the spin-carrying electron position  on one of two monomer sites 
 in a molecular dimer may exist, as is illustrated in Fig.1. Similar glassy dielectric behavior was also observed in EtMe$_3$Sb[Pd(dmit)$_2$]$_2$ \cite{Jawad2}. Strong coupling between the spin and the charge degrees of freedom was also observed in recent microwave measurements of the in-plane dielectric function of $\kappa$-(ET)$_2$Cu$_2$(CN)$_3$ \cite{Poirier}, while its importance to the spin-liquid behavior was pointed out theoretically \cite{Hotta,Dayal}. In fact, there still exists some controversy concerning whether the charge imbalance within dimers really exists \cite{Itoh} or not \cite{Sedlmeier}, and if it exists, on what length and time scales.

\begin{figure}
 \includegraphics[width=8cm]{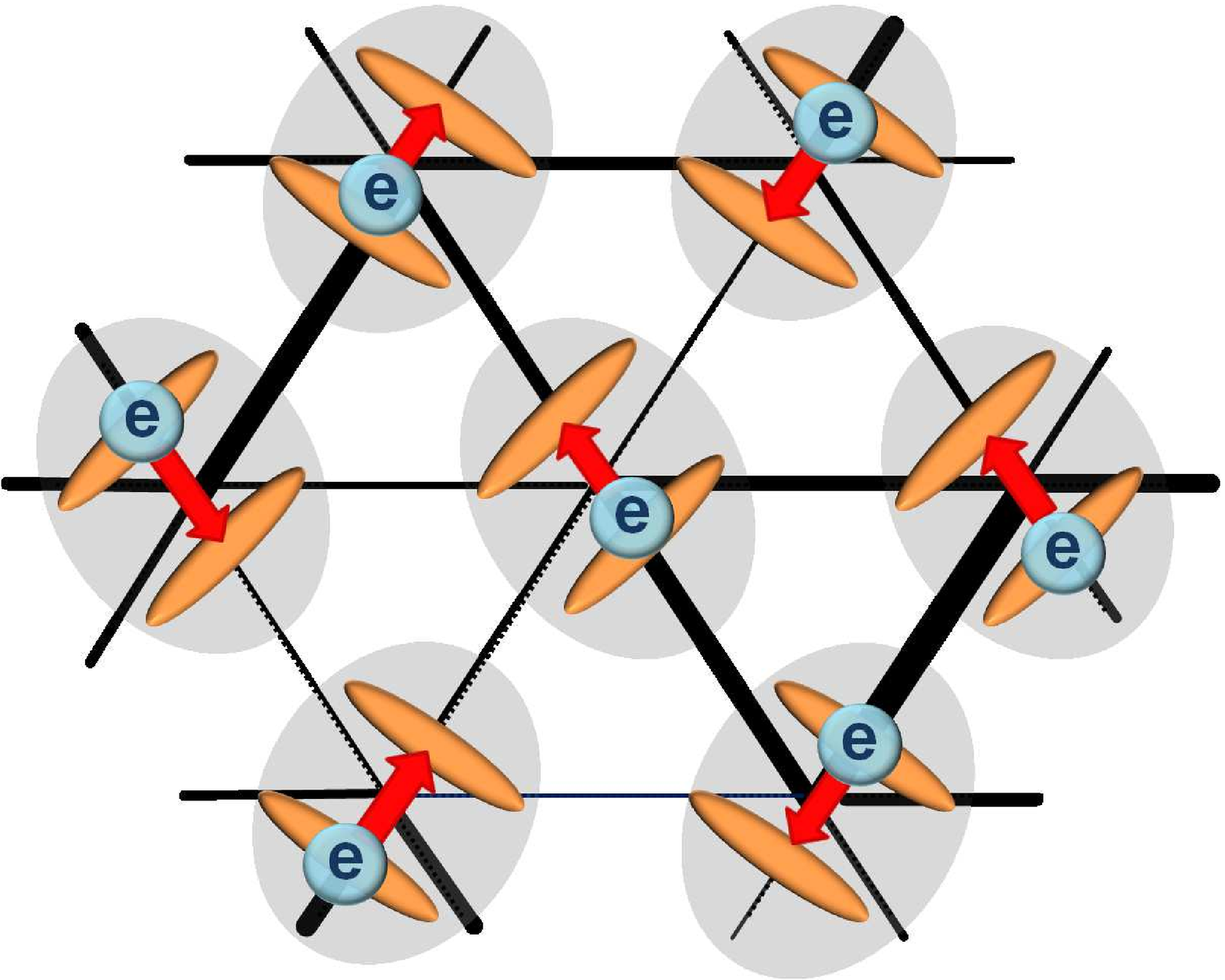}
 \caption{(Color online) Illustration of the random freezing of the electric polarization at each dimer molecule. The glassy dielectric response experimentally observed in some organic salts suggests the biased position of the spin-carrying electron in a molecular dimer occurring at macroscopic time scales. The electric polarization appears at each dimer molecule in a spatially random manner, as indicated by arrows. This might result in the random modification of the exchange couplings acting between two $S=1/2$ spins on neighboring dimer molecules, spanning from a weaker $J$ (thin bond) to a stronger $J$ (thick bond).
}
\end{figure}

 The random freezing of the electric polarization at each triangular site gives rise to the modulation of the exchange constant from bond to bond of the lattice. Since this freezing is almost static at low temperatures, the effective spin Hamiltonian describing the quantum spin liquid might be the bond-random $S=1/2$ Heisenberg model on the triangular lattice with a spatially random exchange coupling, rather than the corresponding regular model. An assumption implicit in this modelling is the separation of two relevant time scales. Namely, fluctuations of the electric polarization are much slower than those of the spins.

 Hence, we study the zero-temperature and finite-temperature properties of the $S=1/2$ triangular-lattice Heisenberg model with the random nearest-neighbor antiferromagnetic coupling, with varying the extent of the quenched randomness.  By analyzing finite-lattice properties of the model by means of an exact-diagonalizaion technique, we find that, when the randomness is sufficiently strong, the gapless spin-liquid state, a state without a magnetic LRO and with a vanishing energy gap, is stabilized at low temperatures. In particular, the $T$-linear specific heat is observed in this quantum spin-liquid regime. Such a randomness-induced quantum spin-liquid state is argued to be the so-called ``random-singlet phase'' \cite{Dasgupta,Bhatt,Fisher,Lin} or the ``valence-bond glass (VBG) phase'' \cite{Tarzia,Singh}, where the static spin singlet is formed in a spatially random manner.

\section{The Model and the method}

 We consider the AF bond-random $S=1/2$ quantum Heisenberg model on the triangular lattice whose Hamiltonian is given by
\begin{equation}
 {\cal H}=\sum_{<ij>} J_{ij}\vec S_i\cdot \vec S_j - H \sum_i S_i^z ,
\end{equation}
where $\vec S_i=(S_i^x,S_i^y,S_i^z)$ is a spin-1/2 operator at the $i$-th site on the triangular lattice, $H$ is the magnetic-field intensity, and $J_{ij}>0$ is the random nearest-neighbor AF coupling obeying the bond-independent uniform distribution between [$(1-\Delta )J$, $(1+\Delta )J$], with the mean $J$. The parameter $\Delta$ represents the extent of the randomness: $\Delta=0$ corresponds to the regular system and $\Delta=1$ to the maximally random system. Although a preliminary calculation of the model was made in Ref.\cite{Imada}, where a small amount of randomness was claimed to induce the diverging susceptibility in the $T\rightarrow 0$ limit, the nature of the low-temperature state of the model has remained largely unknown.

  While the detailed form and the extent of the randomness in the ET salt is not known, a recent theoretical analysis based on the quarter-filled extended Hubbard model suggested that it could be of significant amount \cite{Hotta,Naka}. While some correlations might exist between the bond orientation on the lattice and the magnitude of the exchange coupling $J_{ij}$, we assume here for simplicity that the randomness in $J_{ij}$ obeys a common bond-independent distribution. This is partly because the organic salts at issue are located close to the Mott instability and further-neighbor interactions are probably important, effectively smoothing out the discreteness and  the correlation in randomness, but partly because the detailed form of the randomness is often irrelevant to the fundamental physics of random systems. 

 By means of an exact diagonalization method,  both the zero- and finite-temperature properties of the model are computed for finite lattices. The total number of spins $N$ is $N=$ 9, 12, 15, 18, 21, 24, 27, 30 for $T=0$, and $N=$ 9, 12, 15, 18 for $T>0$, periodic boundary conditions being employed. The maximum size for the chirality-related quantities (see below) are $N=24$. For $N=$ 18 and 24, two different lattice shapes are investigated. Sample average is taken over 500($N=9,12,15$), 250($N=18,21$),160($N=24$), 80($N=27$) and 24($N=30$) independent bond realizations in the $T=0$ calculation, while  500($N=9,12$), 80($N=15$) and 40($N=18$) in the $T>0$ calculation.  In what follows, the energy, temperature and magnetic field are normalized in units of $J$.

\section{Zero-temperature properties}

 We first examine the existence or nonexistence of the magnetic LRO at $T=0$, by computing (i) the squared sublattice magnetization $m_s^2$ associated with the three-sublattice order, and (ii) the spin-glass(SG)-type spin freezing parameter $\bar q$ which takes a nonzero value when the spin is frozen either in a spatially periodic or random manner.

 The squared sublattice magnetization $m_s^2$ is defined by
\begin{eqnarray}
m_s^2 &=& \frac{1}{3} \sum_{\alpha} \left[ \left\langle \frac{1}{(N/6)(N/6+1)} \left( \sum_{i\in \alpha} \vec S_i \right)^2 \right\rangle \right] \nonumber \\ 
      &=& \frac{12}{N(N+6)} \sum_{\alpha} \left[ \sum_{i,j\in \alpha} \left\langle \vec S_i \cdot \vec S_j \right\rangle \right] 
, 
\end{eqnarray}
where $\alpha=1,2,3$ denotes three sublattices of the triangular lattice, and the sum over $i\in \alpha$ ($i,j\in \alpha$) is taken over all sites $i$ ($i$ and $j$) belonging to the $\alpha$-th sublattice. the symbol $\langle \cdots \rangle$ denotes the ground-state expectation value or the thermal average, and $\left[\cdots \right]$ denotes the average over the bond disorder.

 The spin freezing parameter $\bar q$ monitors the spin freezing of any kind, including the spatially random one. It is defined by

\begin{equation}
\bar q =  \frac{1}{N}\sqrt{ \left[ \sum_{i,j} \left\langle  \vec S_i\cdot \vec S_j \right\rangle ^2 \right] } , 
\end{equation}
where the sum over $i$ and $j$ is taken over all sites of the lattice.

\begin{figure*}
 \includegraphics[width=8cm]{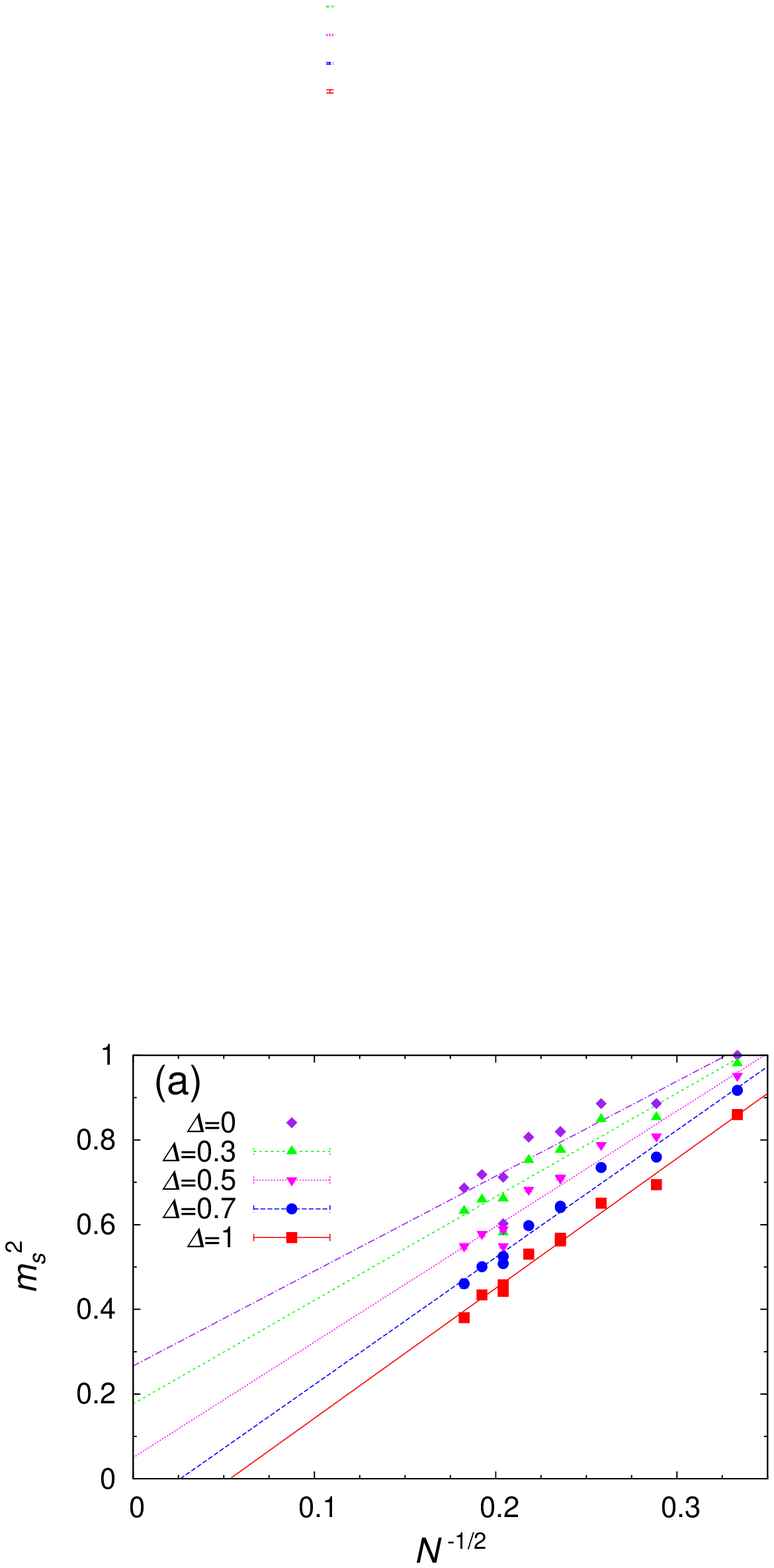}
 \includegraphics[width=8cm]{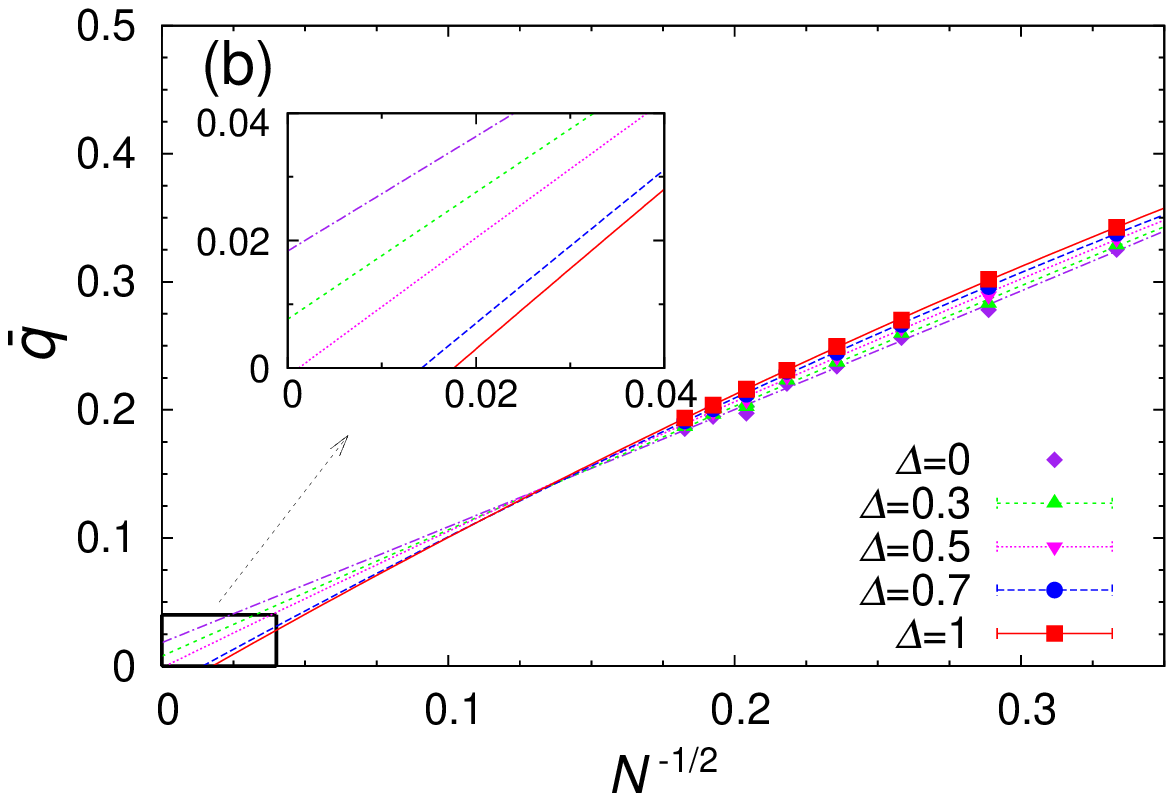}
 \includegraphics[width=8cm]{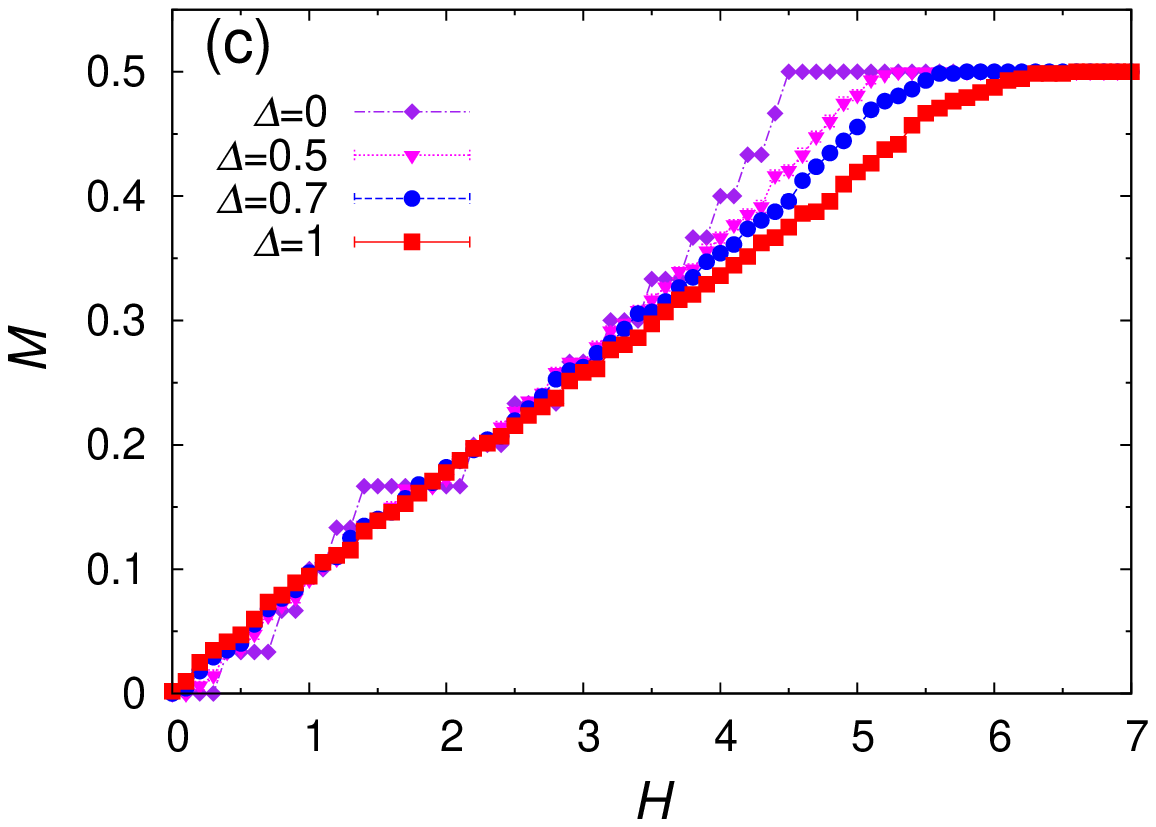}
 \caption{(Color online) Ground-state $T=0$ properties of the model.
 (a) The squared sublattice magnetization $m_s^2$ versus $1/\sqrt{N}$ ($N$ is the total number of spins) for various values of the randomness $\Delta$. The lines are linear fits of the data. (b) The spin freezing parameter $\bar q$ versus $1/\sqrt{N}$ for various values of the randomness $\Delta$. The lines are the quadratic fits based on equation (4). (c) The magnetization curve for various values of the randomness $\Delta$. The lattice size is $N=30$.}
\end{figure*}

 In Fig.2(a), we show the system-size $N$ dependence of the squared sublattice magnetization $m_s^2$ of the ground state for various values of the randomness parameter $\Delta$.  In the regular case $\Delta=0$, previous numerical studies indicated that the model exhibited a nonzero $m_s^2$, $m_s$ being about a half of the saturation value \cite{Bernu,Capriotti}. In the Heisenberg model sustaining a nonzero $m_s^2$, the spin-wave form is expected for its size dependence,
\begin{equation}
 m_s^2 = m_{s,\infty}^2 + \frac{c_1}{\sqrt{N}} + \frac{c_2}{N} + \cdots .
\end{equation}
Indeed, an extrapolation of our present data for $\Delta=0$ based on equation (4) with $c_2=0$ yields a nonzero $m_s^2$-value consistent with the earlier result. A similar extrapolation yields  for $\Delta>0$ a nonzero value for $\Delta \lesssim 0.6$, but a vanishing value for $\Delta \gsim 0.6$, as can be seen from Fig.2(a). This suggests that the three-sublattice N\'eel LRO disappears when the randomness parameter $\Delta$ exceeds a critical value $\Delta_c= 0.6\pm 0.15$. 

 Anyway, the three-sublattice N\'eel LRO goes away for sufficiently strong randomness $\Delta > \Delta_c$.  The next question is whether the disappearance of the N\'eel LRO leads to the SG-type frozen ground state, or to the more exotic spin-liquid-type ground state. In order to examine this point, we plot in Fig.2(b) the spin-freezing parameter $\bar q$ as a function of $1/\sqrt N$. The data turn out to exhibit some curvature here, suggesting the existence of the subleading correction term. Hence, we fit the data to the spin-wave form with the subleading correction, equation (4) with $c_2\neq 0$. For the non-random case $\Delta=0$, the fit yields a nonzero $\bar q\simeq 0.02$. As $\Delta$ grows, the curvature gets less pushing down the extrapolated $\bar q$ value, and for $\Delta \gtrsim 0.5$ the extrapolated $\bar q$-value tends to vanish. The  estimated critical value $\Delta_c=0.5\pm 0.15$ agrees within the error bar with $\Delta_c$ obtained above from the vanishing of $m
 _s^2$.  This coincidence favors a scenario that the three-sublattice N\'eel order gives away to the spin-liquid state which does not sustain any type of magnetic LRO, even of the SG type, when the randomness exceeds a critical value $\Delta_c\simeq 0.6$. 

 In Fig.2(c), we show the ground-state magnetization per spin $M$ for our largest size $N=30$. In the non-random case, the magnetization curve is known to exhibit a plateau at one third of the saturation value, the 1/3 plateau. As the randomness gets stronger, such a plateau goes away. In the spin-liquid regime $\Delta \gtrsim \Delta_c$, the magnetization curve exhibits a near-linear behavior  without any discernible structure for a wide field range.

 We also compute the quantities related to the so-called scalar chirality $\chi_t =\vec S_1\cdot \vec S_2\times \vec S_3$ ($\vec S_1$, $\vec S_2$ and $\vec S_3$ are three corner spins on a triangle $t$), with the knowledge in mind that the chirality sometimes exhibits a strong ordering tendency, even stronger than the spin: under certain conditions, it could lead to the chiral spin-liquid state or to the chiral-glass state \cite{Viet}. We compute both the local scalar-chirality amplitude $\chi_{local}$ and the chiral freezing parameter $\bar q_\chi$. The former is defined by
\begin{equation}
\chi_{local} = \sqrt{ \frac{3}{N} \left[ \sum_{t} \left\langle \chi_t^2 \right\rangle \right] } , 
\end{equation}
where the sum over $t$ is taken over all $N/3$ upward triangles  with a common sublattice arrangement. The chiral freezing parameter is defined by
\begin{equation}
\bar q_\chi = \frac{3}{N}\sqrt{ \left[ \sum_{t,t'} \left\langle  \chi_t \chi_{t'} \right\rangle ^2 \right] } .
\end{equation}
where the sum over $t$ and $t'$ is taken over all $N/3$ upward triangles with a common sublattice arrangement.

The computed $\chi_{local}$ and $\bar q_\chi$ are shown in Fig.3. As can be seen from the inset, the local scalar-chirality amplitude  becomes nonzero for any $\Delta$, even including $\Delta=0$, indicating that the spin tends to take noncoplanar configuration  in the ground state at the local level, {\it i.e.\/}, at each triangle. This is in contrast to the classical case where $\chi_{local}=0$ is expected at least for $\Delta=0$. Obviously, this local non-coplanarity is borne by quantum fluctuations. While the scalar chirality takes a nonzero value locally, Fig.3 indicates that the long-range scalar chirality, both the uniform one and the glass-type random one, tends to zero for $N\rightarrow \infty$ irrespective of the $\Delta$-value. Hence, no chiral spin-liquid phase nor the chiral-glass phase occurs in this quantal model. We also examined a possilbe bond-nematic order by computing the associated bond-nematic freezing parameter, to find that the model does not sustain the bond-nematic order for any value of the randomness parameter $\Delta$.

\begin{figure*}
 \includegraphics[width=8cm]{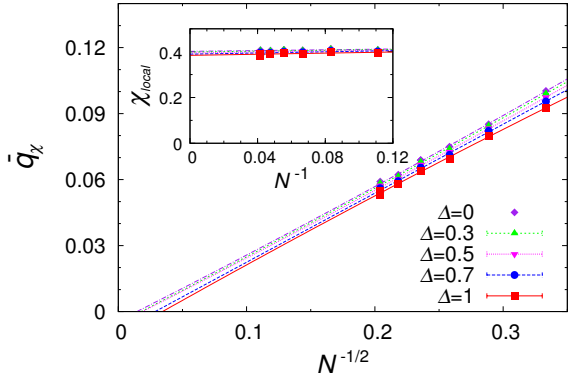}
 \caption{(Color online) The chiral freezing parameter $\bar q_{\chi}$ at $T=0$ is plotted versus $1/\sqrt{N}$ for various values of the randomness $\Delta$. The lines are linear fits of the data. In  the inset, the local scalar-chirality amplitude $\chi_{local}$ at $T=0$ is plotted versus $1/N$ for various values of the randomness $\Delta$. The lines are linear fits of the data.}
\end{figure*}

\section{Finite-temperature properties}

 In this section, we study the finite-temperature properties of the model. In Fig.4(a), we show the temperature dependence of the specific heat per spin $C$ (in units of Boltzmann constant) for the random cases of $\Delta=0.7$ and 1.0. At lower temperatures $T\lsim 0.1$, the specific heat exhibits a $T$-linear behavior $C \simeq \gamma T$, quite different from the behavior of the non-random model. Hence, the spin-liquid phase at $\Delta > \Delta_c$ is characterized by the $T$-linear specific heat, with the $\gamma$-value estimated to be $\gamma \simeq 0.57$ for $\Delta=0.7$. If we use the experimentally estimated $J$-value of $\kappa$-(ET)$_2$Cu$_2$(CN)$_3$ to be $J\sim 250$K \cite{Shimizu}, we get $\gamma \simeq 19$mJK$^{-2}$mol$^{-1}$ which is not far from the value determined from the specific-heat measurements on the ET salt $\gamma \simeq 12 $mJK$^{-2}$mol$^{-1}$ \cite{SYamashita}. As shown in the inset, the specific heat turns out to be insensitive to applied fields wit
 hout any appreciable field dependence up to a field of $0.1J$.  

 In Fig.4(b), we show the temperature dependence of the magnetic susceptibility per spin $\chi$ for several values of $\Delta$.  For smaller $\Delta$, the susceptibility goes to zero in the $T\rightarrow 0$ limit with a finite gap. At $\Delta=0.7$, it tends to a finite value, while at $\Delta=1.0$ it tends to diverge toward $T=0$ obeying the Curie law $\propto 1/T$.  In contrast to the observation of Ref.\cite{Imada}, the Curie-like diverging component arises only when a considerable amount of randomness $\Delta \gtrsim 0.7$ is introduced. The existence of a weak Curie-like component suggests that, in strongly random systems, a small fraction of free spins ($\sim$ 2\%) are generated at low temperatures.

\begin{figure*}
 \includegraphics[width=8cm]{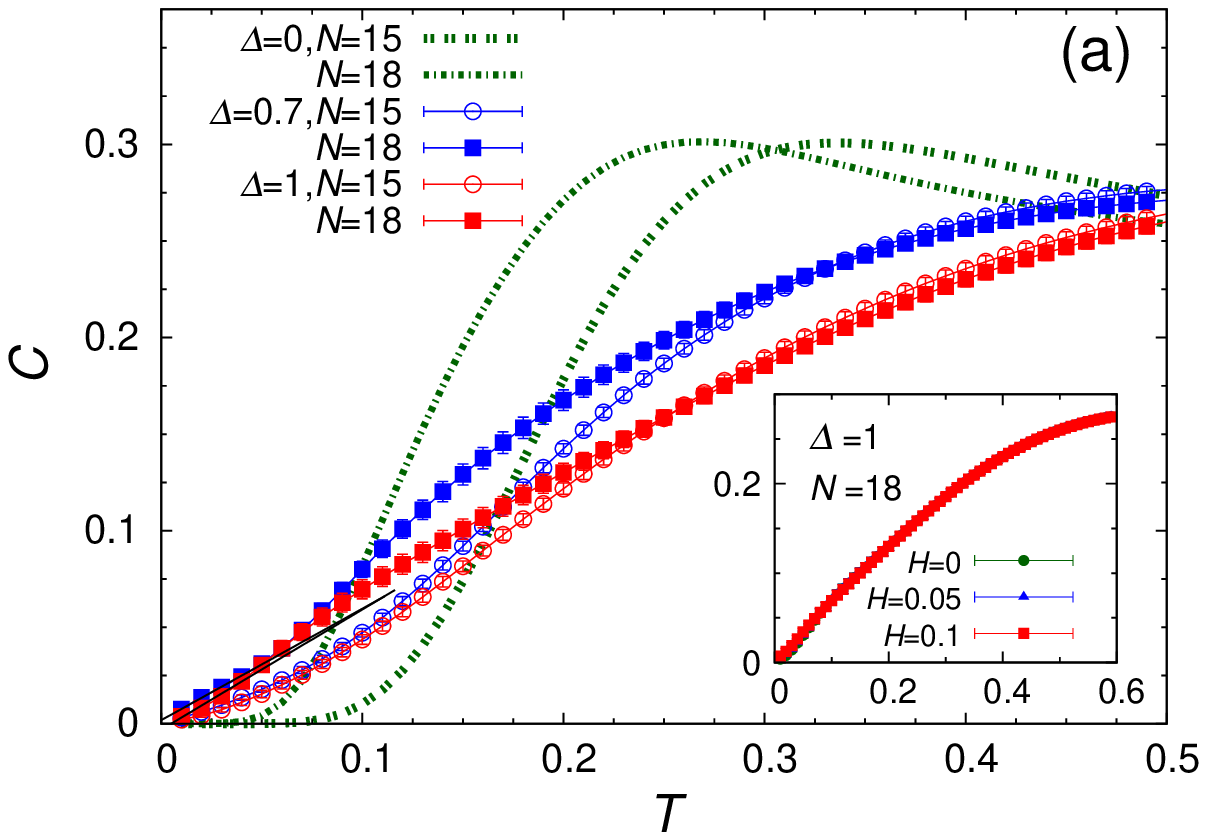}
 \includegraphics[width=8cm]{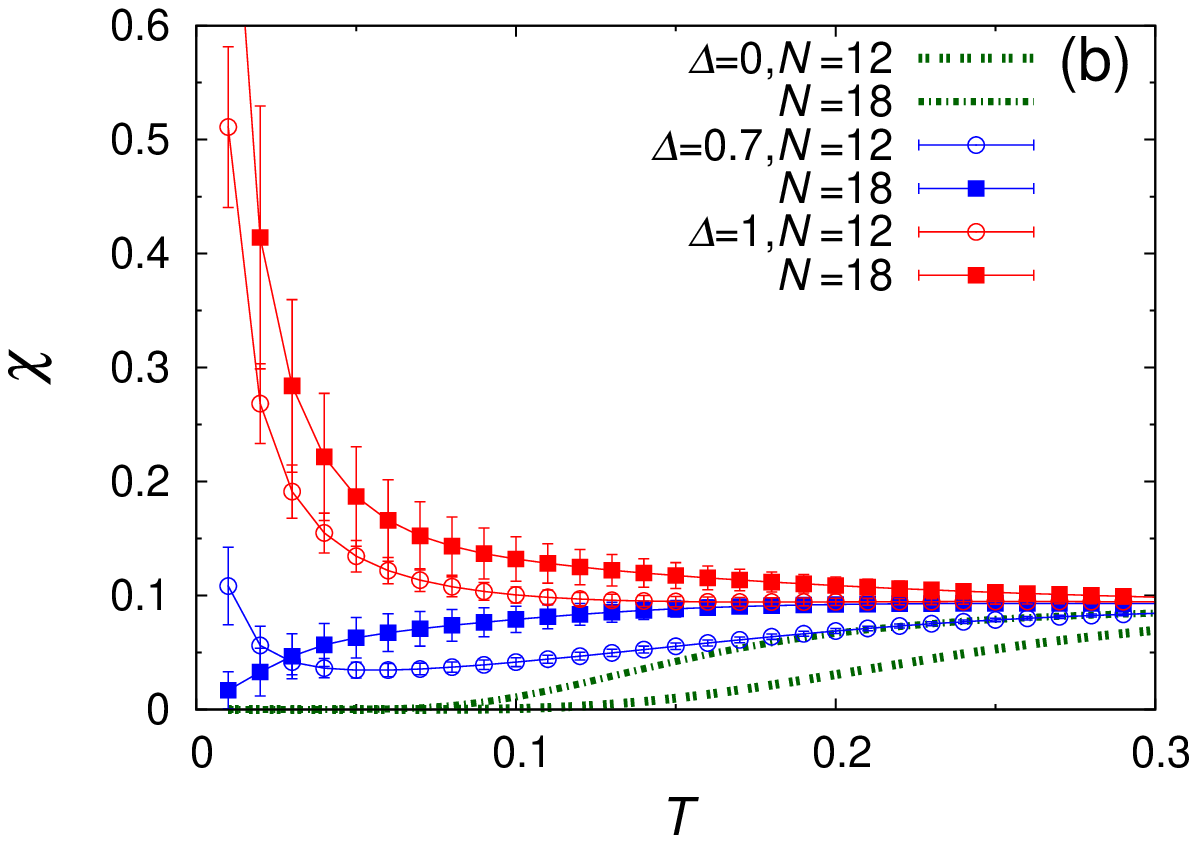}
\caption{
(Color online)  (a) The temperature dependence of the specific heat per spin $C$  for the randomness $\Delta=0,0.7$ and 1.0, and for sizes $N=15$ and 18. The lines are linear fits of the low-temperature data.  The inset represents the corresponding data under magnetic fields for the randomness $\Delta=1.0$ and for the size $N=18$. (b) The temperature dependence of the uniform susceptibility per spin $\chi$ for the randomness $\Delta=0,0.7$ and 1.0, and for the sizes $N=12$ and 18.}
\end{figure*}
 We also compute the NMR relaxation rate $1/T_1$ at finite temperatures \cite{Sakai} according to the relation,

\begin{eqnarray}
\frac{1}{T_1} &=& \lim_{\omega \rightarrow 0} \frac{T}{\omega} \sum_{\bm q}\chi^{''}({\bm q},\omega) , \\ 
\chi^{''}({\bm q},\omega) &=& (1-e^{-\beta \omega}) A \sum_{m,n} e^{-\beta E_m} | \langle m|{\bm S}_{{\bm q}}^z|n\rangle |^2 \delta(\omega-E_m+E_n) / Z,
\end{eqnarray}
where $\beta=1/(k_BT)$ is the inverse temperature, ${\bm S}_{{\bm q}}^z$ is the Fourier transform of the $z$-component of the spin, $Z=\sum_{m}e^{-\beta E_m}$ is the partition function, and $A$ is taken to be a constant, its possible $q$-dependence being neglected. The computation required here is more demanding than that for other quantities due to the off-diagonal matrix elements in eq.(8). Because of this, the number of samples is limited here to 300 ($N=6$ and 12) and 5 ($N=18$), although the error bar is not so large. The delta function is replaced by the Lorentzian with the width $\delta$. The $\delta\rightarrow 0$ and the $\omega \rightarrow 0$ limits  are taken numerically examining the convergence of the results, by making computations for successively smaller $\delta$ and $\omega$. We finally put $\delta=10^{-6}$ and $\omega=10^{-8}$. The constant $A$ is chosen somewhat arbitrary such that the $1/T_1$-value for $\Delta=0$ and $N=12$ becomes unity at $T=1$.

 In Fig.5, we show the temperature dependence of the computed $1/T_1$ for $\Delta=0.7$, which is compared with that of the regular model $\Delta=0$. The data exhibit gapless behaviors characterized by the exponent around $3/2\sim 2$: refer to the lines in the inset. Experimentally, the exponents close to these values are reported, {\it i.e.\/}, $\sim 1.5$ in the ET salt \cite{Shimizu2}, and $\sim 2$ in the dmit salt \cite{Itou2}, whereas the temperature range of the power-law behavior seems lower in experiments. It might also be worth mentioning that our data of $N=12$ and 18 exhibit a broad peak, which is totally absent for the smallest size of $N=6$. The peak tends to become more pronounced for larger sizes, while the peak temperature goes down quickly with increasing $N$. This observation hints that this broad peak of $1/T_1$ might originate from the short-range order effect associated with more than $\sim 10$ spins, not from a dimer nor a single triangle.
  Some cooperative phenomenon such as the $Z_2$-vortex dissociation may be a possible candidate of the observed peak behavior \cite{KawamuraMiyashita}.

\begin{figure*}
 \includegraphics[width=8cm]{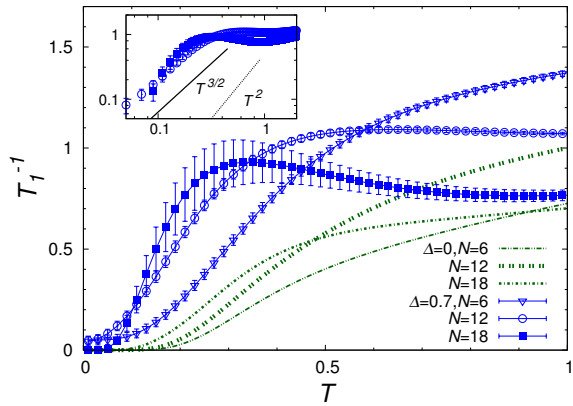}
\caption{ 
(Color online) The temperature dependence of the NMR relaxation rate $1/T_1$ for the randomness $\Delta=0$ and 0.7, and for the sizes $N=6,\ 12$ and 18. The inset represents the corresponding log-log plot for $\Delta=0.7$. The two lines represent the power-law behaviors with the exponents $3/2$ and 2, respectively.
}
\end{figure*}

 In the spin-liquid regime, the ground states of majority of samples with even $N$ turn out to be spin-singlet states corresponding to the total spin $S=0$ (a doublet corresponding to $S=1/2$ for odd $N$). The rate of even-$N$ samples with spin-triplet ground-states, $R$, is shown in the inset of Fig.6. As the randomness gets stronger, $R$ tends to increase, contributing to the diverging susceptibility at $T\rightarrow 0$. We also find that most of the singlet ground states tend to have triplet first excited states. For example, among 153 (143) $N=24$ samples with singlet-ground states, 73\% (81\%) have triplet first excited states for $\Delta=0.7$ ($\Delta=1$). The observation that most of the ground states of even-$N$ samples stay as spin singlets with the total spin $S=0$ even for larger $N$, particularly no sample with $S\geq 2$ found, gives an independent proof that the ground state for $\Delta \geq \Delta_c$ is not the SG state, since the SG ground state entails the tot
 al magnetization increasing proportional to $\sqrt{N}$ for larger $N$, which is clearly inconsistent with our present observation.

 The gapless nature of the ground state can be confirmed by examining the size dependence of the gap energy. In Fig.6, the mean gap energy $\bar \epsilon_g = \left[ \epsilon_g \right]$, {\it i.e.\/}, the energy difference between the ground state and the first excited state averaged over samples, is plotted versus $1/N$ for $\Delta=0.7$ and 1. Linear extrapolation of the data for each even- and odd-$N$ series with a common $\bar \epsilon_g$ value yields $\bar \epsilon_g(N=\infty)=0.01\pm 0.03$ and $0.002\pm 0.02$ for $\Delta=0.7$ and 1, respectively.

 Thus, the $S=1/2$ random AF Heisenberg model with a moderately strong randomness exhibits a spin-liquid ground state, {\it i.e.\/}, a state without the conventional N\'eel ordering nor the spin freezing, accompanied by gapless behaviors including the $T$-linear specific heat.

\begin{figure*}
 \includegraphics[width=8cm]{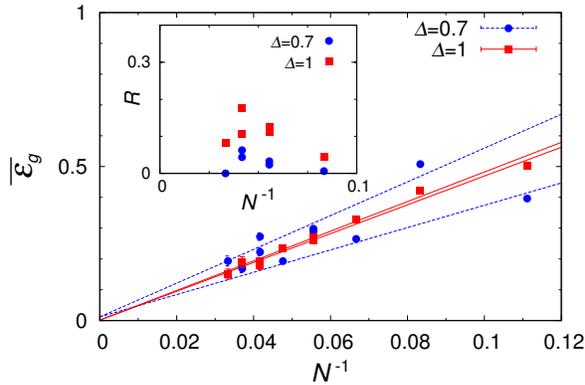}
 \caption{(Color online) The mean gap energy $\bar \epsilon_g$ plotted versus $1/N$ for the randomness $\Delta=0.7$ and 1. The inset exhibits the rate of even-$N$ samples with triplet grounds states.
}
\end{figure*}

\section{Discussion and summary}

 What is the nature of the gapless spin-liquid state as realized in the present random model?  We infer the following picture for the gapless spin-liquid ground state observed here. As the $J$-value is randomly distributed, the binding energy of spin-singlet dimers would also have a wide distribution, a large gap energy for the strong bond with larger $J$, and a small gap energy for the weak bond with smaller $J$. The low-energy excitation of the system then would be the one where the weakest singlet bond is broken turning into a triplet. The state resembles the random-singlet phase \cite{Dasgupta,Bhatt,Fisher,Lin} or the VBG phase \cite{Tarzia,Singh} discussed in the literature. Due to the distribution of $J$ and of the gap energy, a gapless or nearly gapless behavior is generically expected irrespective of the detailed form of the randomness. The appearance of a small fraction of free spins and the resulting Curie-like low-temperature susceptibility is also expected in this
  picture, since the dimer spin pairing might leave some unpaired ``orphan'' spins. 

 Next, we wish to discuss the possible relation of our present results to the recent experiments on the organic salts. We have presumed in our model construction the strong coupling between the electron-polarization and the spin degrees of freedom observed in recent experiments \cite{Jawad,Jawad2,Manna,Poirier}. Then, our finding of the gapless spin-liquid behavior in the spin sector for a moderately strong randomness ($\Delta \gtrsim 0.6$) might provide a useful reference in understanding the experimental results on the organic salts $\kappa$-(ET)$_2$Cu$_2$(CN)$_3$ and EtMe$_3$Sb[Pd(dmit)$_2$]$_2$. Indeed, almost all features of the present model are consistent with the available experimental results. Namely, (i) spins do not freeze down to very low temperature \cite{Shimizu,Ohira,Pratt}. (ii) The low-temperature specific heat exhibits a $T$-linear behavior which is insensitive to magnetic fields, with a coefficient $\gamma$ compatible with the experimental value \cite{SYama
 shita,SYamashita2}. (iii) The susceptibility exhibits a gapless behavior, consistently with experiments \cite{Shimizu,Watanabe}. (iv) The magnetization curve exhibits a linear field dependence for a wide range of applied fields without a discernible anomaly, consistently with the recent experiment \cite{Watanabe}. (v) The randomness-induced spin-liquid state is intrinsically inhomogeneous, which is consistent with the NMR observation for the ET salt of the inhomogeneous internal field induced by applied fields \cite{Shimizu2} (notice that applied fields tend to break the spin singlet, giving rise to an induced moment preferably at weaker $J_{ij}$ bonds). For the dmit salt, the experimental evidence of the inhomogeniety reported so far seems less, although the glassy dielectric response \cite{Jawad2} and the large Curie component in the susceptibility (refer to the caption of Fig.2 of Ref.\cite{Watanabe}) were reported.

  The randomness-induced spin-liquid state of the present model is not an RVB state, but rather a random-singlet state or a VBG state. The gapless or nearly gapless feature is a natural consequence of the inevitable distribution of the binding energy of the spin-singlet dimer, some of which would have a vanishing or very small binding energy. 

 The random-singlet state picture might naturally explain another notable experimental feature that the gapless or nearly gapless behavior is observed quite stably, {\it i.e.\/}, either in $\kappa$-(ET)$_2$Cu$_2$(CN)$_3$ or EtMe$_3$Sb[Pd(dmit)$_2$]$_2$, under deuteration \cite{Watanabe}, and under various pressure \cite{Kurosaki}. As the gapless feature of the random-singlet state arises from the weakly bound spin dimers or from the nearly unpaired spins, such robustness should be a necessary consequence of the spin-singlet dimer formation on the randomly distributed $J_{ij}$, irrespective of the detailed form of the randomness. In particular, the gapless feature as observed here has no direct connection to the quantum criticality. 

 One might ask how frustration is essential in stabilizing the randomness-induced quantum spin-liquid state observed here. To get an insight into this question, we also investigated the corresponding $S=1/2$ bond-random Heisenberg antiferromagnet on the square lattice possessing no frustration. We have then found that, in this unfrustrated model, the AF LRO persists up to the maximal randomness $\Delta=1$. It means that frustration is certainly playing a role in stabilizing the quantum spin-liquid state as observed here. 

 Finally, we wish to emphasize that, although we considered  as the origin of the quenched randomness the glassy dielectric behavior in the present paper, the randomness-induced quantum-spin-liquid behavior could arise quite generically independent of the detailed physical origin of the quenched randomness so long as it is moderately strong. For example, in organic salts, quenched randomness might artificially be introduced by, {\it e.g.\/}, X-ray radiation or rapid quenching. Furthermore, our model is quite generic so that the result could be applicable even to inorganic random magnets. In this connection, it might be interesting to point out that the S=1/2 random distorted-triangular-lattice antiferromagnet CsCu$_2$(Br$_{1-x}$Cl$_{x}$)$_4$ was observed to show a spin-liquid-like behavior for a certain range of $x$ corresponding to strong randomness \cite{Ono}.

 In summary, motivated by the recent experimental observation of the glassy dielectric response, we investigated as a minimal model of organic salts $\kappa$-(ET)$_2$Cu$_2$(CN)$_3$ and EtMe$_3$Sb[Pd(dmit)$_2$]$_2$ the zero- and finite-temperature properties of the spin-1/2 triangular-lattice Heisenberg antiferromagnet with the bond-random interaction. We then found that the model exhibited a gapless quantum-spin-liquid behavior consistently with the recent experimental observations.

\section*{Acknowledgements}
 The authors are thankful to Dr. T. Okubo, Dr. H. Fukuyama, Dr. K. Kanoda, Dr. Y. Matsuda, Dr. Y. Nakazawa, Dr. M. Yamashita, Dr. C. Hotta, Dr. T. Itou, Dr. T. Ono and Dr. H. Tanaka for useful discussion.  The calculation was performed by use of TITPACK Ver.2.  We also thank ISSP, the University of Tokyo for providing us with CPU time. This study is supported by Grants-in-Aid for Scientific Research on Priority Areas ``Novel State of Matter Induced by Frustration'' (No.19052006, 19052008), and by Grants-in-Aid for Scientific Research No.23340109, 23540388, 24540348, 25247064.

%
%\clearpage


\begin{thebibliography}{99}

\bibitem{Levy} B.G. Levi:  Physics Today {\bf 60} (2007) 16.

\bibitem{Lee-review} P.A. Lee: Science {\bf 321}  (2008) 1306.

\bibitem{Balents}  L. Balents: Nature {\bf 464} (2010) 199.

\bibitem{Anderson} P.W. Anderson: Mater. Res. Bull. {\bf 8} (1973) 153.

\bibitem{Shimizu} Y. Shimizu, K. Miyagawa, K. Kanoda, M. Maesato and G. Saito:  Phys. Rev. Lett. {\bf 91} (2003) 107001.

\bibitem{Kurosaki} Y. Kurosaki, Y. Shimizu, K. Miyagawa, K. Kanoda and G. Saito: Phys. Rev. Lett. {\bf 95} (2005) 177001.

\bibitem{Shimizu2} Y. Shimizu, K. Miyagawa, K. Kanoda, M. Maesato and G. Saito:  Phys. Rev. B {\bf 73} (2006) 140407.

\bibitem{Ohira} S. Ohira, Y. Shimizu, K. Kanoda and G. Saito, J. Low Temp. Phys. {\bf 142} (2006) 153.

\bibitem{SYamashita} S. Yamashita, Y. Nakazawa, M. Oguni, Y. Oshima, H. Nojiri, Y. Shimizu, K. Miyagawa and K. Kanoda: Nature Physics {\bf 4} (2008) 459.

\bibitem{MYamashita} M. Yamashita, N. Nakata, Y. Kasahara, T. Sasaki, N. Yoneyama, N. Kobayashi, S. Fujimoto, T. Shibauchi and Y. Matsuda:  Nature Physics {\bf 5} (2009) 44.

\bibitem{Manna} R. S.Manna, M. de Souza, A. Br\"uhl, J.A. Schlueter and M. Lang, Phys. Rev. Lett. {\bf 104} (2010) 016403.

\bibitem{Jawad} M.A. Jawad, I. Terasaki, T. Sasaki, N. Yoneyama, N. Kobayashi, Y. Uesu and C. Hotta:  Phys. Rev. B {\bf 82} (2010) 125119.

\bibitem{Pratt}  F.L. Pratt, P.J. Baker, S.J. Blundell, T. Lancaster, S. Ohira-Kawamura, C. Baines, Y. Shimizu, K. Kanoda, I. watanabe and G. Saito: Nature {\bf 471} (2011) 612.

\bibitem{Poirier} M. Poirier, S. Parent, A. C\^ot\'e, K. Miyagawa, K. Kanoda and Y.Shimizu:  Phys. Rev. B {\bf 85} (2012) 134444.

\bibitem{Sedlmeier} K. Sedlmeier, S. Els\"asser, D. Neubauer, R. Beyer, D. Wu, T. Ivek, S. Tomi\'c, J.A. Schlueter and M. Dressel:  Phys. Rev. B {\bf 86} (2012) 245103.

\bibitem{Itoh} K. Itoh, H. Itoh, M. Naka, S. Saito, I. Hosako, N. Yoneyama, S. Ishihara, T. Sasaki and S. Iwai: Phys. Rev. Lett. {\bf 110} (2013) 106401.

\bibitem{Itou} T. Itou, A. Oyamada, S. Maegawa, M. Tamura and R. Kato:  Phys. Rev. B {\bf 77} (2008) 104413.

\bibitem{Itou2} T. Itou, A. Oyamada, S. Maegawa and R. Kato:  Nature Physics {\bf 6} (2010) 673.

\bibitem{MYamashita2} M. Yamashita, N. Nakata, Y. Senshu, M. Nagata, H.M. Yamamoto, R. Kato, T. Shibauchi and Y. Matsuda: Science {\bf 328}  (2010) 1246.

\bibitem{SYamashita2} S. Yamashita, T. Yamamoto, Y. Nakazawa, M. Tamura and R. Kato: Nature Commun. {\bf 2} (2011) 275.

\bibitem{Watanabe} D. Watanabe, M. Yamashita, S. Tonegawa, Y. Oshima, H.M. Yamamoto, R. Kato, I. Sheikin, K. Behnia, T. Terashima, S. Uji, T. Shibauchi and Y. Matsuda:  Nature Commun. {\bf 3} (2012) 1090.

\bibitem{Jawad2}  M.A. Jawad, N. Tajima, R. Kato, I. Terasaki:  Phys. Rev. B {\bf 88} (2013) 075139.

\bibitem{Bernu} B. Bernu, P. Lecheminant, C. Lhuillier and L. Pierre:  Phys. Rev. B {\bf 50} (1994) 10048.

\bibitem{Capriotti} L. Capriotti, A.E. Trumper and S. Sorella: Phys. Rev. Lett. {\bf 82} (1999) 3899.

\bibitem{LiMing} W. LiMing, G. Misguich, P. Sindzingre and C. Lhullier:  Phys. Rev. B {\bf 62} (2000) 6372.

\bibitem{Morita} H. Morita, S. Watanabe and M. Imada:  J. Phys. Soc. Jpn. {\bf 71} (2002) 2109.

\bibitem{Motrunich} O.I. Motrunich:  Phys. Rev. B {\bf 72} (2005) 045105.

\bibitem{Lee} S.-S. Lee and  P.A. Lee:  Phys. Rev. Lett. {\bf 95} (2005) 036403.

\bibitem{Yunoki} S. Yunoki and S. Sorrela:  Phys. Rev. B {\bf 74} (2006) 014408.

\bibitem{Lee2}  S.-S. Lee, P.A. Lee and T. Senthil: Phys. Rev. Lett. {\bf 98} (2007) 067006.

\bibitem{Xu} Y. Qi, C. Xu and S. Sachdev: Phys. Rev. Lett. {\bf 102} (2009) 176401.

\bibitem{Tocchio} L.F. Tocchio, A. Parola, C. Gros and F.Becca: Phys. Rev. B {\bf 80} (2009) 064419.

\bibitem{Tocchio2} L.F. Tocchio, H. Feldner, F. Becca, R. Valenti and C. Gros: Phys. Rev. B {\bf 87} (2013) 035143.

\bibitem{Hotta} C. Hotta:  Phys. Rev. B {\bf 82} (2010) 241104.

\bibitem{Naka} M. Naka and S Ishihara:  J. Phys. Soc. Jpn. {\bf 79} (2010) 063707.

\bibitem{Dayal} S. Dayal, R.T. Clay, H. Li and S. Mazumdar:  Phys. Rev. B {\bf 83} 245106 (2011).

\bibitem{Dasgupta} C. Dasgupta and S.-K. Ma: Phys. Rev. B {\bf 22} (1980) 1305.

\bibitem{Bhatt} R.N. Bhatt and P.A. Lee: Phys. Rev. Lett. {\bf 48} (1982) 344.

\bibitem{Fisher}  D.S. Fisher: Phys. Rev. B {\bf 50} (1994) 3799.

\bibitem{Lin} Y.-C. Lin, R. M\'elin, H. Rieger and F. Igl\'oi: Phys. Rev. B {\bf 68} (2003) 024424.

\bibitem{Tarzia} M. Tarzia and G. Biroli: Europhys. Lett. {\bf 82} (2008) 67008.

\bibitem{Singh} R.R.P. Singh:  Phys. Rev. Lett. {\bf 104} (2010) 177203.

\bibitem{Imada} M. Imada: J. Phys. Soc. Jpn. {\bf 56} (1987) 881.

\bibitem{Viet} See, {\it e.g.\/}, D.X. Viet, and H. Kawamura: Phys. Rev. Lett. {\bf 102} (2009) 027202.

\bibitem{Sakai} T. Sakai and Y. Takahashi: J. Phys. Soc. Jpn. {\bf 70} (2001) 272.

\bibitem{KawamuraMiyashita} H. Kawamura and S. Miyashita: J. Phys. Soc. Jpn. {\bf 53} (1984) 4138.

\bibitem{Ono} T. Ono, H. Tanaka, T. Nakagomi, O. Kolomiyets, H. Mitamura, F. Ishikawa, T. Goto, K. Nakajima, A. Oosawa, Y. Koike, K. Kakurai, J. Klenke, P. Smeibidle, M. Meissner and H.A. Katori: J. Phys. Soc. Jpn. Suppl. {\bf 74
} (2005) 135.


\end{thebibliography}
\end{document}